\documentclass[12pt]{iopart}

\usepackage{iopams}  
\usepackage{graphicx}
\begin{document}

\title[]{Breaking of parallelograms in presence of torsion: an equivalent alternative approach to detect gravitational waves}

\author{S Nayeh$^1$, A. Latifi$^2$
\footnote{Corresponding autor: A. Latifi,
Department of Physics, Faculty of Science, Qom University of Technology, Qom, Iran}
, S. Arbabi$^2$, M  Ghominejad$^1$}
\address{$^1$ Physics Department, Semnan University, Semnan, Iran }
\address{$^2$ Department of Physics, Faculty of Science, Qom University of Technology, Qom, Iran}
\ead{latifi@qut.ac.ir}

\vspace{10pt}
\begin{indented}
\item[] November 2015
\end{indented}
\vspace{10pt}
\begin{abstract}
The equations for gravitational plane waves produced by a typical binary system as a solution of linear approximation of Einstein equations is derived. The dynamics of the corresponding gravitational field is analyzed in a 4-dimensional space-time manifold, endowed with a metric and taking into account the torsion. In this context, the geometrical reason of the existence of torsion due to the presence of gravitational waves, as an asymmetry of connection coefficients with respect of the swapping of indices's  is highlighted. In a laser interferometer gravitational detector The delay time between the arrivals of the two laser beams traveling back and forth along the two arms of in presence of gravitational waves, is interpreted from this point of view. The geometrical interpretation of torsion, links this delay time to the breaking of the parallelogram formed by the trajectories of the laser beams in space-time. This delay is calculated for a typical NS-NS binary pulsar in two specific orientations with respect of the experimental device, corresponding to different polarizations of the gravitational waves. These values are related to the relative length variation of the detector\rq{}s arms in presence of gravitational waves, and shown to be completely equivalent to the results obtained in the context of the standard General Relativity.
\end{abstract}

%
\vspace{2pc}

\section{Introduction}
\hskip9mm Einstein General Relativity (GR) is an attempt for a full geometrization of physics laws. Hence, in presence of a gravitational field, the space-time is curved. The gravitational interaction lets (spinless) particles to follow geodesics of the curved space-time. However  the most general Lorentz connection has two fundamental properties: curvature and torsion \cite{ko-no}. Cartan was the first who asked the question: ``Why should matter produce only curvature?'' As a possible answer the Einstein-Cartan theory \cite{sabata} has been formulated in which the Christoffel connection is replaced by a more general connection including both curvature and torsion.\\

This formulation at the microscopic level takes into account the fact that the matter is represented by elementary particles characterized by their mass and their spins. Adopting the geometrical point of view of GR, mass produces curvature of space-time while spin is source of torsion. 
At macroscopic level, where spin vanishes, Einstein-Cartan theory coincides with GR. However no experimental evidence allows to justify Einstein-Cartan theory. Moreover, in the interaction of the electromagnetic field with gravitation, Einstein-Cartan theory violates the $U(1)$ gauge invariance of Maxwell theory \cite{hehl-maccrea}.\\

An other attempt to include torsion in gravitation, is the Teleparallel Gravity, where the Christoffel connection is replaced by the so called Weitzenb\"ock connection \cite{arcos-pereira}. In contrast to Christoffel connection, the Weitzenb\"ock connection has a non-vanishing  torsion but vanishing curvature. \\

In the present study, in the context of an alternative extension of GR, we consider a weak gravitational field corresponding to a region of space with almost no curvature (our solar system) and we show how the disturbance created by gravitational waves (GW) produces torsion.\\

A 4-dimensional space-time manifold, endowed with a metric and taking into account the torsion, bring us out of the Riemann space and will produce the \lq\lq{}breaking of parallelograms\rq\rq{}. We relate a quantitative evaluation of this phenomenon to the relative change of length of a laser interferometer GW detector\rq{}s arms in presence of GW produced by a binary system. We show that this relative change of length is completely equivalent, in first order, to the results obtained by the standard GR.\\

For the numerical application, we consider a typical NS-NS binary pulsar (1913 +16) with two equal masses  $M=1.4M_\odot=2.77\times10^{30}\hskip1mm kg$, on an almost circular orbit with radius $a=3\times10^7\hskip1mm {m}$ and at the distance from Earth $z_0=500 \hskip1mm pc = 1.54 \times 10^{19}\hskip1mm m$.

\section{Weak field metric, linear approximation of Einstein equations}
\label{sec:2}
\hskip9mm  Weak gravitational field corresponds to a region of space-time which is almost ``flat''. This means that we assume a background with Minkowskian structure in space-time with metric $\eta_{\mu\nu}$. Accordingly, throughout such region, there exist coordinate systems $x^\mu$ in which the space-time metric takes the form 
\begin{equation}
g_{\mu\nu}=\eta_{\mu\nu}+h_{\mu\nu} \quad\quad |h_{\mu\nu}|\ll 1, \label{eq1}
\end{equation}
and the first and higher partial derivatives of $h_{\mu\nu}$ are also small. This interpretation is consistent with GR, as well as with the point of view of field theory, according to which a field propagates on a background space-time \cite{blanchet}.\\

Note that for a Riemann manifold $h_{\mu\nu}$ must be symmetric with respect of the swapping of its indices. But in presence of torsion, neither $h_{\mu\nu}$ nor $g_{\mu\nu}$ will be symmetric. We will come back to this point in Sect.~\ref{sec:5}.\\

To be able to describe a time-varying weak gravitational field, we have to assume 
\begin{equation}
\partial_0\hskip1mm g_{\mu\nu}\neq0\quad\quad \partial_0\hskip1mm h_{\mu\nu}\neq0.
 \label{eq2}
 \end{equation}
Given a Lorentz frame, $e^a= e^a_\mu dx^\mu$, an infinitesimal general coordinate transformation takes the form 
\begin{equation}
x'^\mu=x^\mu+\xi^\mu(x),
\label{eq3}\end{equation}
where the $\xi^\mu(x)$ are four arbitrary functions of position of the same order of smallness as the $h_{\mu\nu}$. From (\ref{eq3}) we have
\begin{equation}
\frac{\partial x'^\mu}{\partial x^\nu}=\delta^\mu_\nu +\partial_\nu \xi^\mu.
\label{eq4}
\end{equation}
Thus, at first order of small quantities, the metric transforms as
\begin{equation}
g'_{\mu\nu}=\frac{\partial x^\rho}{\partial x'^\mu}\frac{\partial x^\sigma}{\partial x'^\nu}g_{\rho\sigma}=\eta_{\mu\nu}+h_{\mu\nu}-\partial_\mu \xi_\nu-\partial_\nu \xi_\mu, \label{eq5}
\end{equation}
where $\xi_\mu=\eta_{\mu\nu}\xi^\nu$. The linearized form of the Ricci tensor $R_{\mu\nu}$ and the Ricci scalar $R$ are 
\begin{equation}
R_{\mu\nu}=\frac12 \left( \partial_\nu \partial_\mu h+{\opensquare}^2 h_{\mu\nu}
-\partial_\nu \partial_\rho h_\mu^\rho-\partial_\rho \partial_\mu h_\nu^\rho \right)
, \label{eq6}
\end{equation}
with $h\equiv h^\sigma_\sigma$\hskip1mm, \hskip3mm $\opensquare^{2}\equiv \partial_\sigma\partial^\sigma$ and $ R=\opensquare^2 h_{\mu\nu}-\partial_\rho \partial_\mu h^{\nu\rho}$. \\
\vskip1mm\hskip1mm
By defining $\bar{h}_{\mu\nu}$, the ``trace reverse'' of $h_{\mu\nu}$ 
\begin{equation}
\bar{h}_{\mu\nu}={h}_{\mu\nu}-\frac{1}{2}\eta_{\mu\nu}h
, \label{eq8}
\end{equation}
the Einstein equations 
\begin{equation}
R_{\mu\nu}-\frac{1}{2}g_{\mu\nu}R=-\kappa T_{\mu\nu}
, \label{eq9}
\end{equation}
become \cite{hobson}
\begin{equation}
\opensquare^2 \bar{h}_{\mu\nu}+\eta_{\mu\nu}\partial_\rho\partial_\sigma\bar{h}^{\rho\sigma}-
\partial_\mu\partial_\rho\bar{h}^{\rho}_\nu-
\partial_\nu\partial_\rho\bar{h}^{\rho}_\mu=-2\kappa T_{\mu\nu}. \label{eq10}
\end{equation}
Note that in this linearized form, $h_{\mu\nu}$ contains all the informations about the gravity. These equations can be simplified further by making use of the gauge transformation
\begin{equation}
{h'}_{\mu\nu}={h}_{\mu\nu}-\partial_\mu\xi_\nu-\partial_\nu\xi_\mu, \label{eq11}
\end{equation}
where the components of the trace reverse transform of $h'_{\mu\nu}$ are defined as follows 
\begin{equation}
\bar{h'}^{\mu\nu}={h'}^{\mu\nu}-\frac{1}{2}\eta^{\mu\nu}h'
. \label{eq12}
\end{equation}
By substituting (\ref{eq11}) in  (\ref{eq12}) and differentiating, we have 
\begin{equation}
\partial_\rho\bar{h'}^{\mu\nu}=\partial_\rho\bar{h}^{\mu\nu}-\opensquare^2\xi^\mu
. \label{eq13}
\end{equation}
The most convenient choice in this class of coordinate systems is to choose the function $\xi^\mu(x)$ satisfying the condition $\opensquare^2\xi^\mu=\partial_\rho\bar{h}^{\mu\rho}$. By dropping the primes, the linearized field equation in the new gauge becomes 
\begin{equation}
\opensquare^2\bar{h}_{\mu\nu}=-2\kappa T_{\mu\nu}
, \label{eq14}
\end{equation}
provided $\bar{h}^{\mu\nu}$ satisfy the gauge condition
\begin{equation}
\partial_\mu\bar{h}^{\mu\nu}=0.
 \label{eq15}
\end{equation}
Finally in vacuo, the linearized Einstein equations under the  gauge conditions, reads \cite{Aldrovandi}
\begin{equation}
\opensquare^2\bar{h}^{\mu\nu}=0
. \label{eq16}
\end{equation}
It worth remarking that the nonlinearity of the original Einstein equations comes from the fact that any energy-momentum acts as a source for gravitational fields, including the energy-momentum associated with the gravitational field itself. By linearizing the field equation, we ignore this fact.
\section{Gravitational plane waves: solution of linearized Einstein equations in vacuo }
\label{sec:3}
\hskip9mm 
A plane-wave solution of the relativistic wave equation (\ref{eq16}) has the form 
\begin{equation}
\bar{h}^{\mu\nu}=A^{\mu\nu}\exp\left[ik_\rho x^\rho\right]
, \label{eq17}
\end{equation}
where $A^{\mu\nu}$ are constants and the wave vectors $k_\rho$ satisfy
\begin{equation}
k_\rho k^\rho=0
. \label{eq18}
\end{equation}
On the other hand, the gauge condition implies
\begin{equation}
A^{\mu\nu}k_\nu=0
. \label{eq19}
\end{equation}
The physical solution corresponds to the real part of the gravitational plane-wave namely
\begin{equation}
\bar{h}^{\mu\nu}=\frac{1}{2}A^{\mu\nu}\exp\left[ik_\rho x^\rho\right]+\frac{1}{2}\left(A^{\mu\nu}\right)^*\exp\left[-ik_\rho x^\rho\right]
, \label{eq20}
\end{equation}
which clearly is the superposition of two plane-waves. For a wave traveling in the $x^3$-direction and making use of the Lorentz gauge transformation, satisfying the condition  (\ref{eq15}), known as the transverse-traceless gauge \cite{Aldrovandi}, we have
\begin{equation}
    \label{eq21}
{{A}^{\mu \nu }}=\left[
\begin{array}{cccc}
   0 & 0 & 0 & 0 \\
   0 & a & b & 0 \\
   0 & b & -a & 0 \\
   0 & 0 & 0 & 0 \\
   \end{array}
   \right] \hskip1mm .
\end{equation}
\vskip1mm \hskip1mm 
By introducing the linear polarization tensors $e^{\mu\nu}_1$ and $e^{\mu\nu}_2$, where the component are obtained by setting respectively  ($a=1, \hskip2mm  b=0$) and ($a=0, \hskip2mm b=1$) in (\ref{eq21}), ${A}^{\mu \nu }$ can be written as the following linear combination
\begin{equation}
A^{\mu\nu}=ae^{\mu\nu}_1+be^{\mu\nu}_2. \label{eq22}
\end{equation}
\section{Generation of GW by a binary system}
\label{sec:4}
\begin{figure}
\centering
  \includegraphics[width=0.5\textwidth]{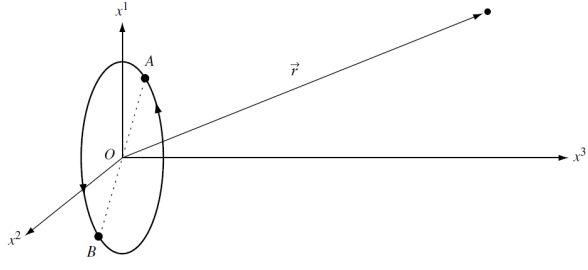}
\caption{Two particles with equal masses $M$ rotate in circular orbit with radius $a$ about their center of mass.}
\label{fig:1}       
\end{figure}

\hskip9mm 
By considering the matter distribution of the binary system localized near the origin $O$ of our coordinate system and our field point $\overrightarrow{r}$ at the distance $r$ from $O$  (Fig.~\ref{fig:1}) and for $r$ large compared to the spatial extent of the source, we may use the compact source approximation \cite{hobson}
\begin{equation}
\bar{h}^{\mu\nu}(ct, \overrightarrow{r})=-\frac{4G}{c^4r}\int {T^{\mu\nu}(ct-r,\overrightarrow{y})d^3y }\hskip1mm .\label{eq23}
\end{equation}
By taking our spatial coordinates $x^i$ to correspond to the ``center of momentum'' frame of the source, we have $\rho^i=0$ ($i \neq 0$). Thus
\begin{equation}
\bar{h}^{00}=-\frac{4GM}{c^2r}, \quad \quad\quad\quad \bar{h}^{0i}=\bar{h}^{i0}=0. \label{eq24}
\end{equation}
The remaining components are
\begin{equation}
\bar{h}^{ij}(ct, \overrightarrow{r})=-\frac{4G}{c^4r}\int {T^{ij}(ct-r,\overrightarrow{y})d^3y }.\label{eq25}
\end{equation}
Standard calculations give the quadrupole formula
\begin{equation}
\bar{h}^{ij}\left( ct,\overrightarrow{r} \right)=-\frac{2G}{c^6r}\left[ \frac{d^2I^{ij}(ct')}{dt'^2} \right]_{ct'=ct-r}, \label{eq26}
\end{equation}
where we define the quadrupole-moment tensor of the energy density of the source 
\begin{equation}
I^{ij}(ct)=\int {T^{00}(ct,\overrightarrow{r})y^iy^jd^3y }=\int {\rho (ct,\overrightarrow{r})y^iy^jd^3y }. \label{eq27}
\end{equation}
Notice that $I^{ij}$ are constant tensors on each hyper-surface of constant time.\\

In the particular case of a binary system of equal masses $M$ moving non-relativistically in a circular orbit of radius $a$ and an angular speed $\Omega$, the coordinates of $A$ and $B$ are
\begin{equation}
\eqalign{[x^i_A]=\left(a\cos\Omega t, a\sin\Omega t, 0\right) \cr
[x^i_B]=-\left(a\cos\Omega t, a\sin\Omega t, 0\right)}. \label{eq28}
\end{equation}
Treating the motion in the Newtonian limit requires
\begin{equation}
\Omega=\sqrt{\frac{GM}{4a^3}}
. \label{eq29}
\end{equation}
Thus the proper density of the system $A-B$ as a source of GW is given by
\begin{eqnarray}
\fl\nonumber \rho \left( ct,\overrightarrow{r} \right)=\\
\fl M\left[ \delta \left( {{x}^{1}}-a\cos \Omega t \right)\delta \left( {{x}^{2}}-a\sin \Omega t \right) 
+\delta \left( {{x}^{1}}+a\cos \Omega t \right)\delta \left( {{x}^{2}}+a\sin \Omega t \right) \right]\delta \left( {{x}^{3}} \right). \label{eq30}
\end{eqnarray}
\vskip1mm \hskip1mm
We are going to consider two particular positions of the observer: a) the observer is located on the $x^3$-axis, and b) the observer is located on the $x^1$-axis. \\

  In the case that the observer is located on the $x^3$-axis (at large distance from $O$) the quadrupole-moment tensor (\ref{eq27}) becomes ($T^{00}=\rho$)
\begin{equation}
{{I}^{ij}}\left( ct \right)=M{{c}^{2}}{{a}^{2}}\left[ \begin{array}{ccc}
   1+\cos 2\Omega t & \sin 2\Omega t & 0  \\
   \sin 2\Omega t & 1-\cos 2\Omega t & 0  \\
   0 & 0 & 0  
\end{array} \right]. \label{eq31}
\end{equation}
Using the quadrupole formula (\ref{eq26}) we have
\begin{equation}
{{\bar{h}}^{\mu \nu }}(ct,\overrightarrow{r})=\frac{8GM{{a}^{2}}{{\Omega }^{2}}}{{{c}^{4}}r}\left[ \begin{array}{cccc}
   0 & 0 & 0 & 0  \\
   0 & \cos 2\Omega \left( t-\frac{r}{c} \right) & \sin 2\Omega \left( t-\frac{r}{c} \right) & 0  \\
   0 & \sin 2\Omega \left( t-\frac{r}{c} \right) & -\cos 2\Omega \left( t-\frac{r}{c} \right) & 0  \\
   0 & 0 & 0 & 0  \\
\end{array} \right]. \label{eq32}
\end{equation}
According to (\ref{eq1}) we can write
\begin{equation}
 \fl{g}_{\mu \nu }(ct,\overrightarrow{r})=\left[ \begin{array}{cccc}
   -1 & 0 & 0 & 0  \\
   0 & 1+\frac{8GM{{a}^{2}}{{\Omega }^{2}}}{{{c}^{4}}r}\cos 2\Omega \left( t-\frac{r}{c} \right) & \frac{8GM{{a}^{2}}{{\Omega }^{2}}}{{{c}^{4}}r}\sin 2\Omega \left( t-\frac{r}{c} \right) & 0  \\
   0 &\frac{8GM{{a}^{2}}{{\Omega }^{2}}}{{{c}^{4}}r} \sin 2\Omega \left( t-\frac{r}{c} \right) & 1-\frac{8GM{{a}^{2}}{{\Omega }^{2}}}{{{c}^{4}}r}\cos 2\Omega \left( t-\frac{r}{c} \right) & 0  \\
   0 & 0 & 0 & 1  \\
\end{array} \right]. \label{eq33}
\end{equation}
By setting 
\begin{equation}
h_{+}=\frac{8GM{{a}^{2}}{{\Omega }^{2}}}{{{c}^{4}}r}\cos 2\Omega \left( t-\frac{r}{c} \right) \quad \quad h_{\times}=\frac{8GM{{a}^{2}}{{\Omega }^{2}}}{{{c}^{4}}r}\sin 2\Omega \left( t-\frac{r}{c} \right), \label{eq34}
\end{equation}
we obtain
\begin{equation}
ds^2=-c^2dt^2+ \left( 1+h_{+}\right)(dx^1)^2+\left( 1-h_{+}\right)(dx^2)^2+ (dx^3)^2+2h_{\times}dx^1dx^2. \label{eq35}
\end{equation}
\vskip1mm 
The second position of the observer considered here, is the case the observer is located on the $x^1$-axis. The $\bar{h}^{\mu\nu}$ matrix becomes then
\begin{equation}
{{\bar{h}}^{\mu \nu }}(ct,\overrightarrow{r})=\frac{4GM{{a}^{2}}{{\Omega }^{2}}}{{{c}^{4}}r}\left[ \begin{array}{cccc}
   0 & 0 & 0 & 0  \\
   0 & 0 & 0 & 0  \\
   0 & 0 & -\cos 2\Omega \left( t-\frac{r}{c} \right) & 0  \\
   0 & 0 & 0 &\cos 2\Omega \left( t-\frac{r}{c} \right)  \\
\end{array} \right].  \label{eq36}
\end{equation}
It follows 
\begin{equation}
\fl{g}_{\mu \nu }(ct,\overrightarrow{r})=\left[ \begin{array}{cccc}
   -1 & 0 & 0 & 0  \\
   0 & 1 & 0 & 0  \\
   0  & 0 & 1-\frac{4GM{{a}^{2}}{{\Omega }^{2}}}{{{c}^{4}}r}\cos 2\Omega \left( t-\frac{r}{c} \right) & 0  \\
   0 & 0 & 0 & 1+\frac{4GM{{a}^{2}}{{\Omega }^{2}}}{{{c}^{4}}r}\cos 2\Omega \left( t-\frac{r}{c} \right)  \\
\end{array} \right], \label{37} 
\end{equation}
and in this case the metric reads
\begin{equation}
ds^2=-c^2dt^2+ (dx^1)^2+\left( 1-\frac{h_{+}}{2}\right)(dx^2)^2+\left( 1+\frac{h_{+}}{2}\right) (dx^3)^2.\label{38}
\end{equation}

\section{Gravitational waves and torsion}
\label{sec:5}
\hskip9mm 
In a 4-dimensional space-time manifold endowed with metric and torsion \cite{hehl-vdh}, the torsion tensor $T^\rho_{\mu\nu}$ can be defined as
\begin{equation}
T^\rho_{\mu\nu}=\Gamma^\rho_{\mu\nu}-\Gamma^\rho_{\nu\mu}, \label{eq39}
\end{equation}
where  $\Gamma^\rho_{\mu\nu}$ are the affine connection coefficients. In Einstein GR, it is postulated that $T^\rho_{\mu\nu}=0$ which means that the connection coefficients $\Gamma^\rho_{\mu\nu}$ are chosen to be symmetric with respect of the swapping of indices. \\

Let us examine the reason why the affine connection coefficients $\Gamma^\rho_{\mu\nu}$ could be chosen not to be symmetric. In a 4-dimensional space-time manifold, consider a curvilinear system of coordinates and in this system consider a point. To this point, we associate a rectilinear system of coordinates $\overrightarrow{e}_\mu$ where the vectors $x^\mu$ are  parallel to the tangents of curvilinear system of coordinates, namely
\begin{equation}
\overrightarrow{e}_1(x^1, 0, 0, 0),\quad\overrightarrow{e}_2(0, x^2, 0, 0),\quad\overrightarrow{e}_3(0, 0, x^3, 0),\quad\overrightarrow{e}_4(0, 0, 0, x^4).
 \label{eq40}
\end{equation}
Let us call this rectilinear system, the natural system of coordinates associated to the point $x^\mu$. The natural system will change if we make an infinitesimal displacement $\overrightarrow{ds}$ 
\begin{equation}
\overrightarrow{ds}=\overrightarrow{e_\mu}dx^\mu. \label{eq41}
\end{equation}
The coefficients connecting two natural systems before and after the displacement $\overrightarrow{ds}$ are called the affine connection coefficients $\Gamma^\rho_{\mu\nu}$. Subsequently, we have
 \begin{equation}
\partial_\nu\overrightarrow{e_\mu}=\Gamma^\rho_{\mu\nu}\overrightarrow{e_\rho}. \label{eq42}
\end{equation} 
 On the other hand, an infinitesimal displacement $\overrightarrow{ds}$ can also  be expressed as follows
  \begin{equation}
\overrightarrow{ds}=\frac{\overrightarrow{\partial s}}{\partial x^\mu}dx^\mu. \label{eq43}
\end{equation}
 Now comparing (\ref{eq41}) and (\ref{eq43}), we can write
  \begin{equation}
\frac{\overrightarrow{\partial s}}{\partial x^\mu}=\overrightarrow{e_\mu}. \label{eq44}
\end{equation}
Knowing that in the natural system, which is nothing but an Euclidean space, we have
  \begin{equation}
\frac{\partial}{\partial x^\nu}\left(\frac{\overrightarrow{\partial s}}{\partial x^\mu}\right)=\frac{\partial}{\partial x^\mu}\left(\frac{\overrightarrow{\partial s}}{\partial x^\nu}\right), \label{eq45}
\end{equation}
and the relation (\ref{eq44}) gives
\begin{equation}
\partial_\nu \overrightarrow{e_\mu}=\partial_\mu \overrightarrow{e_\nu}. \label{eq46}
\end{equation}
 Thus (\ref{eq46}) and (\ref{eq42}) lead to the symmetry property of the affine connection coefficients 
 \begin{equation}
 \Gamma^\rho_{\mu\nu}=\Gamma^\rho_{\nu\mu},\label{eq47}
 \end{equation}
 and of course, the lack of this symmetry property :
 \begin{equation}
\partial_\nu \overrightarrow{e_\mu}\neq\partial_\mu \overrightarrow{e_\nu}, \label{eq48}
\end{equation}
 means that a unit vector does not change in a similar way in different direction of the space. In particular the length of $\overrightarrow{e_\mu}$ can be different at different points of space. This bring us out of Riemann manifold which is a metric space defined by a symmetric metric tensor $g_{\mu\nu}$.\\
 
  In this sense, in presence of GW, where the length unit could be different in different directions of space, torsion is required for a complete theory of gravitation.\\
  
   An extreme point of view which ignores the geometrical approach of the GR and According to the Teleparallel Gravity, a contortion plays the role of a gravitational force, similarly to the Lorentz force of electrodynamics \cite{hammond}. This contortion of the space-time could appear in presence of GW \cite{maluf}.
\section{Breaking of parallelograms in presence of torsion}
 \label{sec:6}
\begin{figure}
\centering
  \includegraphics[width=0.5\textwidth]{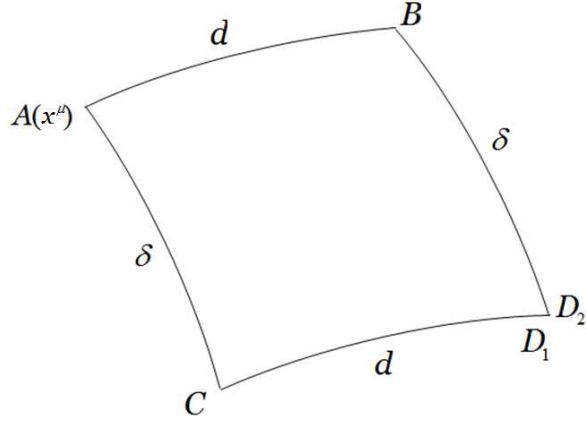}
\caption{Two different displacements $\delta$ and $d$ are applied to the point $A$ along the geodesic $AC$ and $AB$ respectively. A displacement $\delta$ applied to $B$ and a displacement $d$ applied to  $C$ make a quasi-parallelogram in a n-dimensional Riemann space where $D_1$ and $D_2$ coincide.}
\label{fig:2}       
\end{figure}

 \hskip9mm 
 Let us have a glance on how a parallelogram is defined in an n-dimensional Riemann space and how the breaking of parallelograms occur in presence of torsion.\\
 We consider a point $A(x^\mu)$ in an n-dimensional Riemann space. From this point, we can make two different infinitesimal displacements along two different geodesics $AB$ and $AC$ and call them respectively $\delta$ and $d$ (Fig.~\ref{fig:2}).
At the second stage, a $\delta$-displacement is applied to $B$ and a $d$-displacement is applied to $C$. The resulting points, are respectively $D_1$ and $D_2$. In an ordinary differential algebra, we have
 \begin{equation}
 d(\delta x^\mu)=\delta(dx^\mu),\label{eq49}
 \end{equation}
and consequently, the points $D_1$ and $D_2$ coincide. This define a parallelogram or better speaking a quasi-parallelogram in a n-dimensional Riemann space-time. \\

\begin{figure}
\centering
  \includegraphics[width=0.3\textwidth]{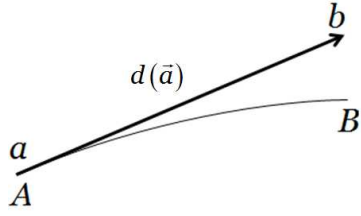}
  \caption{The vector ${ab}$ in the tangent space is obtain by the application of the infinitesimal displacement $d$ to the point $a$.}
\label{fig:3}      
\end{figure}

Now, consider the parallel transport (coordinate invariant transformation) of the set of n vectors $\overrightarrow{e_\mu}$\hskip1mm , defining the reference system at $A$. The transportation along the path $ABD$, transforms $\overrightarrow{e_\mu}$ on a new reference system $(\overrightarrow{e_\mu})_1$\hskip1mm , and the transportation along the path $ACD$, transforms $\overrightarrow{e_\mu}$ on an other set of reference vectors $(\overrightarrow{e_\mu})_2$. Then, let us consider the Euclidean tangent space at $A$ and call all the points of this tangent  space by small letters. Hence, the point $A$ in the Riemann space coincide with the point $a$ of the tangent space. The parallel transport of  $\overrightarrow{\displaystyle e_\mu}$ along $AB$ gives, in the \textit{tangent space}, a new set of reference  vectors $\overrightarrow{e_\mu}+d\overrightarrow{e_\mu}$ at the point $b$ (the counterpart of the point $B$ in the Riemann space). The vector $\overrightarrow{AB}$ in the Riemann space corresponds then, to the vector  $\overrightarrow{ab}$ in the tangent space (Fig.~\ref{fig:3}) and we have
\begin{equation}
\overrightarrow{ab}=\overrightarrow{d(a)}=\overrightarrow{e_\mu}\hskip1mm dx^\mu, \label{eq50}
\end{equation}
and
\begin{equation}
d\overrightarrow{e_\mu}=\Gamma^\rho_{\mu\nu}dx^\nu \overrightarrow{e_\rho}. \label{eq51}
\end{equation}
In the same manner, the displacement $\delta$ of $\overrightarrow{e_\mu}+d\overrightarrow{e_\mu}$ from the point $b$ defines 
\begin{equation}
\overrightarrow{bd_1}=\delta (\overrightarrow{a}+\overrightarrow{d(a)})= \overrightarrow{\delta (a)}+\overrightarrow{\delta d(a)}. \label{eq52}
\end{equation}
So 
\begin{equation}
\overrightarrow{ad_1}=\overrightarrow{ab}+\overrightarrow{bd_1}=\overrightarrow{d(a)}+\overrightarrow{\delta (a)}+\overrightarrow{\delta d
(a)}
. \label{eq53}
\end{equation}
Hence, the reference system $(\overrightarrow{e_\mu})_1$ at $d_1$ reads
\begin{equation}
(\overrightarrow{e_\mu})_1=(\overrightarrow{e_\mu}+d\overrightarrow{e_\mu}) + \delta (\overrightarrow{e_\mu}+d\overrightarrow{e_\mu})=\overrightarrow{e_\mu}+d(\overrightarrow{e_\mu})+\delta (\overrightarrow{e_\mu})+\delta d(\overrightarrow{e_\mu}). \label{eq54}
\end{equation}
The same operation along the path $ACD$ gives
\begin{equation}
\overrightarrow{ad_2}=\overrightarrow{\delta (a)}+\overrightarrow{d(a)}+\overrightarrow{d\delta(a)}
, \label{eq55}
\end{equation}
\begin{equation}
(\overrightarrow{e_\mu})_2=\overrightarrow{e_\mu}+\delta(\overrightarrow{e_\mu})+d(\overrightarrow{e_\mu})+d\delta (\overrightarrow{e_\mu}). \label{eq56}
\end{equation}
So, the vector $\overrightarrow{d_1d_2}$ can be expressed as follows
\begin{eqnarray}
\nonumber \overrightarrow{d_1d_2}&=d\delta (\overrightarrow{a})-\delta d(\overrightarrow{a})\\
\nonumber &=d(\overrightarrow{e}_\mu\delta x^\mu)-\delta(\overrightarrow{e}_\mu\delta x^\mu)\\
&=\overrightarrow{e}_\mu(d\delta x^\mu-\delta dx^\mu)+d\overrightarrow{e_\mu}\delta x^\mu-\delta\overrightarrow{e_\mu} dx^\mu
. \label{eq57}
\end{eqnarray}
Since in tangent Euclidean space 
$
 d(\delta x^\mu)=\delta(dx^\mu),
$
(\ref{eq57}) becomes 
\begin{equation}
\overrightarrow{d_1d_2}=d\overrightarrow{e_\mu}\delta x^\mu-\delta\overrightarrow{e_\mu} dx^\mu
, \label{eq58}
\end{equation}
and using (\ref{eq51}), we obtain
\begin{equation}
\overrightarrow{d_1d_2}=\overrightarrow{e_\rho}\hskip1mm (\Gamma^\rho_{\mu\nu}-\Gamma^\rho_{\nu\mu})dx^\nu\delta x^\mu. \label{eq59}
\end{equation}
\vskip1mm \hskip1mm 
Thus, we can see that in presence of torsion where $\Gamma^\rho_{\mu\nu}\neq\Gamma^\rho_{\nu\mu}$, the quasi-parallelogram $ABCD$ is no more represented by a closed quadrilateral in the Euclidean tangent space. This is one calls the \lq\lq{}breaking of parallelogram\rq\rq{} \cite{Eisenhart}.
\section {Detection of GW produced by a binary system in a laser interferometer GW detector through the breaking of parallelogram}
\label{sect:7}

\begin{figure}
\centering
  \includegraphics[width=0.5\textwidth]{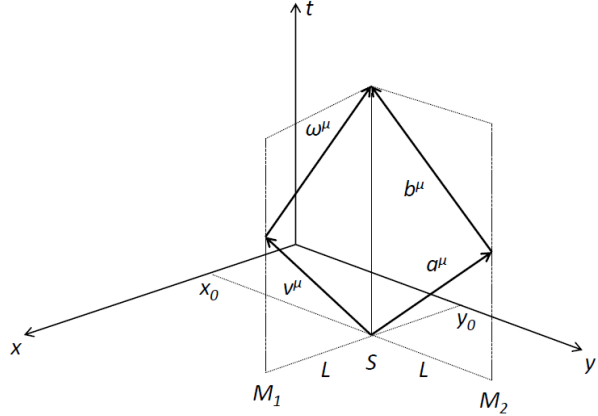}
\caption{Mirrors $M_1$ and $M_2$ are fixed at the end of the arms $SM_1$ and $SM_2$. The light beams go from $S$ to $M_1$ and $M_2$ and after reflection return back to $S$. This trajectories form a parallelogram in the (2+1) dimension space-time.}
\label{fig:4}    
\end{figure}

 \hskip9mm A laser interferometer gravitational waves detector (LIGO \cite{ligo} in USA, VIRGO \cite{virgo} and GEO \cite{geo} in Europe, TAMA \cite{tama} in Japan) consists on laser beams traveling back and forth along perpendicular arms in the $xy$-plane in presence of a plane-fronted GW traveling in the $z$ direction \cite{hough}. The mirrors are fixed at the end of the arms. At a point $S$, the laser beam splits in two. One of the beams travels along the arm of the proper length $L$ in $x$-direction and the other, along the other arm  in $y$-direction with the same  proper length $L$. At the end of each arm, the mirrors $M_1$ and $M_2$ reflect the beams back to the point $S$ (Fig.~\ref{fig:4}).
The traveling beams from $S$ to $M$ and from $M$ back to $S$ form respectively, the vectors $v^\mu$ and $w^\mu$. The beam following the path $SM_2$ and $M_2S$ form the vectors $a^\mu$ and $b^\mu$.\\

\begin{figure}
\centering
  \includegraphics[width=0.8\textwidth]{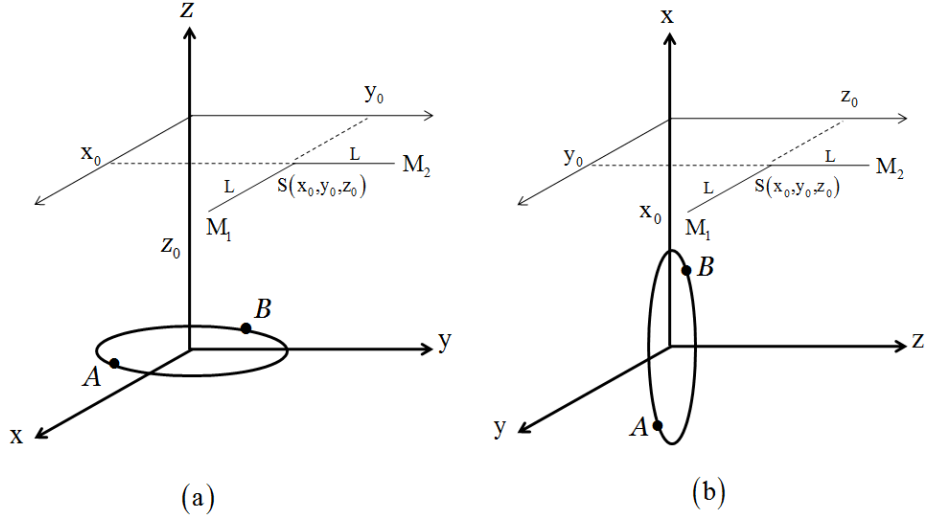}
\caption{ (a) The plane of rotation of the binary system is parallel the $xy$-plane containing the arms of the detector and perpendicular to the $z$-direction. (b) The plane containing the arms of the detector, is parallel to the $yz$-plane.}
\label{fig:5}      
\end{figure}

We are going to consider a binary system, as the source of GW in two different positions with respect to the detector\rq{}s arms (Fig.~\ref{fig:5}).
  \\

In the first position, where the plane of rotation of the binary system is parallel to $xy$-plane containing the arms of the detector, the corresponding metric is the one given in (\ref{eq35}). In this case $v^{\mu}(ct,x,y,z)=(v^0,v^1,0,0)$ satisfying the condition $g_{\mu\nu}V^{\mu}V^{\nu}=0$. For the laser beam traveling in the positive direction along the $x$-axis, the latter condition yields to the following relation
\begin{equation}
g_{00}(v^0)^2+g_{11}(v^1)^2=0. \label{eq60}
\end{equation}
We require from $v^1$ to have the length $L$. Hence, (\ref{eq60}) and (\ref{eq35}) give
\begin{equation}
v^0=\sqrt{-\frac{g_{11}}{g_{00}}}v^1=\sqrt{1+h_{+}}L  \qquad  h_{+}=h_{+}(x_0+L, y_0, z_0, t_0+dt), \label{eq61}
\end{equation}
where $t_0$ is the time coordinate of the emitted light at point $S$, and $t_0+dt$ is the time coordinate of the traveling light of point $M_1$. Therefore the vector $v^\mu$ is defined as follows
\begin{equation}
v^\mu=\left( \sqrt{1+h_{+}}L, L, 0, 0\right)  \quad\quad\quad h_{+}=h_{+}(x_0+L, y_0, z_0, t_0+dt). \label{eq62}
\end{equation}
On a similar way, we define $w^\mu$, $a^\mu$ and $b^\mu$ as follows
\begin{eqnarray}
\nonumber& w^\mu=\left( \sqrt{1+h_{+}}L, -L, 0, 0\right)  \quad\quad\quad h_{+}=h_{+}(x_0, y_0, z_0, t_0+2dt),\\
\nonumber& a^\mu=\left( \sqrt{1-h_{+}}L, 0, L, 0\right)  \quad\quad\quad h_{+}=h_{+}(x_0+L, y_0, z_0, t_0+dt),\\
& b^\mu=\left( \sqrt{1-h_{+}}L, 0, -L, 0\right)  \quad\quad\quad h_{+}=h_{+}(x_0, y_0, z_0, t_0+2dt). \label{eq63}
\end{eqnarray}\vskip5mm \hskip9mm
If we quantify the breaking of the parallelogram in space-time by the difference
\begin{equation}
\Delta^\mu= (v^\mu+w^\mu)-(a^\mu+b^\mu),\label{eq64}
\end{equation}
then, clearly for $i=1, 2, 3$ we have $\Delta^i=0$, and the only nonzero component of $\Delta^\mu$ is $\Delta^0$
\begin{eqnarray}
\fl \nonumber \Delta^0 =& L\{ \sqrt{1+h_{+}(x_0+L, y_0, z_0, t_0+dt)}+\sqrt{1+h_{+}(x_0, y_0, z_0, t_0+2dt)}\\
\fl &-\sqrt{1-h_{+}(x_0, y_0+L, z_0, t_0+dt)}
-\sqrt{1-h_{+}(x_0, y_0, z_0, t_0+2dt)}\}. \label{eq65}
\end{eqnarray}
Obviously we have $dt \ll t_0$, and $L \ll x_0, \hskip 2mm L \ll y_0$. Moreover, knowing that $dt\ll L/c$ and $x_0=ct_0$, one shows easily
\begin{equation}
\frac{dt}{t_0}=\frac{L}{x_0}. \label{eq66}
\end{equation}
The relation (\ref{eq66}) shows that ${dt}/{t_0}$ and ${L}/{x_0}$ are of the same order of smallness. So we can expand $h_{+}$ in terms of $L$ and $dt$, according to 
\begin{eqnarray}
\fl \nonumber & h_{+}(x_0+L, y_0, z_0, t_0+dt)\approx h_{+}(x_0, y_0, z_0, t_0)+ \left(\frac{\partial h_{+}}{\partial x} \right)_{x=x_0}L+\left(\frac{\partial h_{+}}{\partial t} \right)_{t=t_0}dt, \\
\fl \nonumber & h_{+}(x_0, y_0+L, z_0, t_0+dt)\approx h_{+}(x_0, y_0, z_0, t_0)+ \left(\frac{\partial h_{+}}{\partial y} \right)_{y=y_0}L+\left(\frac{\partial h_{+}}{\partial t} \right)_{t=t_0}dt, \\
\fl  & h_{+}(x_0, y_0, z_0, t_0+2dt)\approx h_{+}(x_0, y_0, z_0, t_0)+ \left(\frac{\partial h_{+}}{\partial t} \right)_{t=t_0}2dt.
\label{eq67}
\end{eqnarray}

 On the other hand, $h_{+}$ is sufficiently small (\ref{eq34}) for applying the approximation
\begin{equation}
 \sqrt{1+h_{+}}=1+\frac{1}{2}h_{+},\qquad \quad \sqrt{1-h_{+}}=1-\frac{1}{2}h_{+}. \label{eq68}
\end{equation}
By substituting (\ref{eq67}) and (\ref{eq68}) in (\ref{eq65}), we have
\begin{equation}
\fl\Delta^0= \frac{L}{2}\left( 4 h_{+}(x_0, y_0, z_0, t_0)+\left(\frac{\partial h_{+}}{\partial x} \right)_{x=x_0}L+\left(\frac{\partial h_{+}}{\partial y} \right)_{y=y_0}L+6 \left(\frac{\partial h_{+}}{\partial t} \right)_{t=t_0}dt \right).
 \label{eq69}
\end{equation}
Replacing (\ref{eq34}) (the expression of $h_{+}$) in (\ref{eq69}) and using  $dt_0= L/c$, (\ref{eq69}) reads
\begin{eqnarray}
\fl\nonumber &\Delta^0= \frac{L^2}{2} \left(\frac{8GMa^2\Omega^2}{c^4r_0}\right) \Bigg\{ \frac{4}{L} \cos 2\Omega\left(t_0-\frac{r_0}{c}\right)+ \\
\fl &\frac{x_0+y_0}{r_0}\left(\frac{2\Omega}{c}\sin 2\Omega\left(t_0-\frac{r_0}{c}\right) 
 -\frac{1}{r_0}\cos 2\Omega\left(t_0-\frac{r_0}{c}\right) \right) -\frac{12\Omega}{c}\sin 2\Omega\left(t_0-\frac{r_0}{c}\right)\Bigg\}
\label{eq70}
\end{eqnarray}
As mentioned in Sect.~\ref{sec:2}, we assume an almost Minkowskian structure for the space-time. Therefore, with a good approximation we can write $r_0=\sqrt{x_0^2+y_0^2+z_0^2}$. By setting $x_0=y_0=0$, we have $r_0=z_0$, and
\begin{equation}
 \Delta^0= \frac{L^2}{2} \left(\frac{8GMa^2\Omega^2}{c^4z_0}\right) \Bigg\{ \frac{4}{L} \cos 2\Omega\left(t_0-\frac{z_0}{c}\right) -\frac{12\Omega}{c}\sin 2\Omega\left(t_0-\frac{r_0}{c}\right)\Bigg\}.
\label{eq71}
\end{equation}\vskip1mm\hskip1mm
For a typical NS-NS binary pulsar (1913 + 16), we have  two equal masses  $M=1.4M_\odot=2.77\times10^{30}\hskip1mm kg$ and the distance from Earth $z_0=500 \hskip1mm pc = 1.54 \times 10^{19}\hskip1mm m$ \cite{ju}. By considering the length $L$ of LIGO's arms to be 4 km, we see that  $4/L\gg 12\Omega/c$. Thus, we can neglect the second term in (\ref{eq71}), which yields
\begin{equation}
\Delta^0\simeq\frac{16GMa^2\Omega^2L}{c^4z_0}\cos 2\Omega\left(t_0-\frac{z_0}{c}\right).
\label{eq72}
\end{equation}
The frequency of this system is $f=10^{-3.7}\simeq 2\times 10^{-4}\hskip1mm s^{-1}$\cite{verbunt}. Then according to (\ref{eq29}), we obtain the radius of its circular orbit $a=2.7\times10^7\hskip1mm {m}$.
The delay time between the arrivals of the splitted beams along the two arms of the detector is given by $\Delta\tau=\Delta^0/c$, which gives
\begin{equation}
\Delta\tau\simeq\frac{16GMa^2\Omega^2L}{c^5z_0}\cos 2\Omega\left(t_0-\frac{z_0}{c}\right),
\label{eq73}
\end{equation}
and by replacing all the values, we obtain
 \begin{equation} \left |\Delta\tau\right |\lesssim4.73\times10^{-26}\hskip1mm s \hskip1mm.
 \end{equation}\vskip1mm \hskip1mm
The usual evaluation of the gravitational wave\rq{}s effect on the detctor\rq{}s arms is done through the ratio $\Delta L/L$ where $L$ is the length of the arms in absence of gravitational wave. To compare our result with this usual evaluation, we simply have to notice that $ \Delta L =\Delta^0/2$. Then using the numerical values mentioned above, we obtain
\begin{equation}
\left|\frac{\Delta L}{L}\right|\lesssim\frac{8GMa^2\Omega^2L}{c^5z_0}\simeq 1.77\times 10^{-21}\hskip1mm .
\end{equation}
According to (\ref{eq34}), this means
\begin{equation}
\left|\frac{\Delta L}{L}\right|\lesssim h_{+} \hskip3mm \textsf{and} \hskip3mm \log\left|\frac{\Delta L}{L}\right|\lesssim \log h_{+}\simeq-20.8 \label{final}
\end{equation}
 \vskip1mm
 \hskip1mm Now, let us consider the second position of the binary system with respect of the experimental device. Namely, when the plane formed by the arms of detector is parallel to $yz$-plane (Fig.~\ref{fig:5}). In this case,  $v^{\mu}(ct,x,y,z)=(v^0,0,v^2,0)$ which satisfies the condition $g_{\mu\nu}V^{\mu}V_{\nu}=0$. Again, for a laser beam traveling in the positive direction along the $x$-axis, the latter condition gives
 \begin{equation}
g_{00}(v^0)^2+g_{22}(v^2)^2=0, \label{eq74}
\end{equation}
and requiring $v^2$ to have the length $L$, (\ref{eq74}) and (\ref{eq35}) yield
\begin{equation}
v^0=\sqrt{-\frac{g_{22}}{g_{00}}}v^2=\sqrt{1-\frac{h_{+}}{2}}L \qquad h_{+}=h_{+}(x_0, y_0+L, z_0, t_0+dt). \label{eq75}
\end{equation}
It this situation the vectors $v^\mu$, $w^\mu$, $a^\mu$ and $b^\mu$ are defined as follow
\begin{eqnarray}
\nonumber v^\mu&=\left( \sqrt{1-\frac{h_{+}}{2}}L, 0, L, 0\right)  \qquad h_{+}=h_{+}(x_0, y_0+L, z_0, t_0+dt),\\
\nonumber w^\mu&=\left( \sqrt{1-\frac{h_{+}}{2}}L, 0, -L, 0\right),  \qquad h_{+}=h_{+}(x_0, y_0, z_0, t_0+2dt)\\
\nonumber a^\mu&=\left( \sqrt{1+\frac{h_{+}}{2}}L, 0, 0, L\right),  \qquad h_{+}=h_{+}(x_0, y_0, z_0+L, t_0+dt)\\
 b^\mu&=\left( \sqrt{1+\frac{h_{+}}{2}}L, 0, 0, -L\right)  \qquad h_{+}=h_{+}(x_0, y_0, z_0, t_0+2dt).  \label{eq76}
\end{eqnarray} 
Again, as previously, we expand $h_{+}$ in terms of $L$ and $dt$
\begin{eqnarray}
\fl\nonumber & h_{+}(x_0, y_0+L, z_0, t_0+dt)\approx h_{+}(x_0, y_0, z_0, t_0)+ \left(\frac{\partial h_{+}}{\partial y} \right)_{y=y_0}L+\left(\frac{\partial h_{+}}{\partial t} \right)_{t=t_0}dt, \\
\fl\nonumber & h_{+}(x_0, y_0, z_0+L, t_0+dt)\approx h_{+}(x_0, y_0, z_0, t_0)+ \left(\frac{\partial h_{+}}{\partial z} \right)_{z=z_0}L+\left(\frac{\partial h_{+}}{\partial t} \right)_{t=t_0}dt, \\
\fl &h_{+}(x_0, y_0, z_0, t_0+2dt)\approx h_{+}(x_0, y_0, z_0, t_0)+ \left(\frac{\partial h_{+}}{\partial t} \right)_{t=t_0}2dt,
\end{eqnarray}\label{eq77}
and use the approximation 
\begin{equation}
 \sqrt{1+\frac{h_{+}}{2}}=1+\frac{1}{4}h_{+}\quad \quad \sqrt{1-\frac{h_{+}}{2}}=1-\frac{1}{4}h_{+}. \label{eq78}
\end{equation}
Now using (\ref{eq77}) and (\ref{eq78}), the expression of $\Delta^0$, i.e. (\ref{eq65}) becomes
\begin{equation}
\fl\Delta^0=-\frac{L}{4}\left[ 4 h_{+}(x_0, y_0, z_0, t_0)+\left(\frac{\partial h_{+}}{\partial z} \right)_{z=z_0}L+\left(\frac{\partial h_{+}}{\partial y} \right)_{y=y_0}L+6 \left(\frac{\partial h_{+}}{\partial t} \right)_{t=t_0}dt \right]\label{eq79}
\end{equation}
and using the expression of $h_{+}$ (\ref{eq34}) and taking into account $dt_0= L/c$, we have
 \begin{eqnarray}
\fl\nonumber &\Delta^0=- \frac{L^2}{4} \left(\frac{8GMa^2\Omega^2}{c^4r_0}\right) \Bigg\{ \frac{4}{L} \cos 2\Omega\left(t_0-\frac{r_0}{c}\right)\\
\fl& +\frac{z_0+y_0}{r_0}\left(\frac{2\Omega}{c}\sin 2\Omega\left(t_0-\frac{r_0}{c}\right)-\frac{1}{r_0}\cos 2\Omega\left(t_0-\frac{r_0}{c}\right)\right)
 -\frac{12\Omega}{c}\sin 2\Omega\left(t_0-\frac{r_0}{c}\right)\Bigg\}. \label{eq80}
\end{eqnarray}
In this situation,  $y_0=z_0=0$ and, neglecting the second term in (\ref{eq80}) (since $4/L\gg 12\Omega/c$), $\Delta^0$ becomes
\begin{equation}
\Delta^0\simeq-\frac{8GMa^2\Omega^2L}{c^4x_0}\cos 2\Omega\left(t_0-\frac{x_0}{c}\right). \label{eq81}
\end{equation}
According to our previous definition of delay time $\Delta\tau=\Delta^0/c$, 
\begin{equation}
\Delta\tau\simeq-\frac{8GMa^2\Omega^2L}{c^5x_0}\cos 2\Omega\left(t_0-\frac{x_0}{c}\right).
\label{eq82}
\end{equation}
The typical values used previously give here  
\begin{equation}\left|\Delta\tau\right|\lesssim2.36\times10^{-26}\hskip1mm s\end{equation}
and
\begin{equation}
\left|\frac{\Delta L}{L}\right|\lesssim1.27\hskip1mm \times 10^{-21}\hskip1mm .
\end{equation}
On other words
\begin{equation}
\left|\frac{\Delta L}{L}\right|\lesssim\frac{1}{2}\left|h_{+} \right|\hskip3mm\textsf{and} \hskip3mm \log\left|\frac{\Delta L}{L}\right|\lesssim \log \frac{1}{2}\left|h_{+}\right|\simeq-21.1
\end{equation}
\section{Comparison with results obtained in the context of the standard GR}\label{sect:8}\hskip9mm
Here we recall briefly the method used in the context of the standard GR to evaluate the relative change on the arms length $\Delta L/L$ of a laser-interferometer GW detector.\\

Let us consider two nearby observers freely falling in the field of a weak and plane GW. This wave produces small variations in the proper distance between the two observers A and B. We call the frame attached to one of this observers, let say A (the basic observer), the proper reference frame which is formed by a small Cartesian latticework of measuring rods and synchronized clocks. The space-time coordinates in this frame $(c\hat{t},\hat{x}^1,\hat{x}^2,\hat{x}^3)$ are locally Lorentzian along the whole geodesic of the observer A. Thus the line element of the metric in these coordinates has the form
\begin{equation}
ds^2=-c^2d\hat{t}^2+\delta_{ij}d\hat{x}^i d\hat{x}^j+O\left((\hat{x}^i)^2\right)d\hat{x}^\alpha d\hat{x}^\beta\hskip1mm.\label{metric}
\end{equation}
Notice that this line element deviates from the line element of the flat Minkowski space-time by terms that are at least quadratic in the values of $\hat{x}^i$ .\\

Defining the deviation vector $\hat{\xi}^\alpha$ describing the instantaneous relative position  of the observer B with respect to observer A:
\begin{equation}
\hat{\xi}^\alpha=\left( 0, \hat{x}^1(\hat{t}),\hat{x}^2(\hat{t}),\hat{x}^3(\hat{t})\right)\hskip1mm .\label{distance}
\end{equation}
The relative acceleration ${D^2\hat{\xi}^\alpha}/{d\hat{t}^2}$ is related to the sapce-time curvature through the equation of geodesic deviation \cite{misner}
\begin{equation}
\frac{D^2\hat{\xi}^\alpha}{d\hat{t}^2}=-c^2\hat{R}^\alpha_{\beta\gamma\delta}\hat{u}^\beta\hat{\xi}^\gamma\hat{u}^\delta\hskip1mm ,\label{diff}
\end{equation}
where $\hat{R}^\alpha_{\beta\gamma\delta}$ is the Riemann curvature tensor and $\hat{u}^\beta$ is the 4-velocity of the observer A.\\

The equation (\ref{metric}) implies that the Christoffel Symbols $\hat{\Gamma}^\alpha_{\beta\gamma}$ vanish along the basic geodesic. Taking into account the fact that $\hat{u}^\alpha=(1,0,0,0)$ and using (\ref{distance}) and (\ref{diff}) one obtains
\begin{equation}
\frac{d^2\hat{x}^i}{d\hat{t}^2}=-c^2\hskip1mm \hat{R}_{i0j0}\hat{x}^j+O\left((\hat{x}^i)^3\right)\hskip1mm .\label{diff2}
\end{equation}\vskip1mm
\hskip1mm If one chooses the Traceless Transformed (TT) coordinates (introduced in section 3) in such way that the 4-velocity field needed to define TT coordinates, coincides with the 4-velocity of our basic observer A, then
\begin{equation}
\hat{R}_{i0j0}=R^{TT}_{i0j0}+O(h^2)\hskip1mm .
\end{equation}
For a recent thorough analysis and a proof of the equivalence between the TT and the free falling frame see \cite{rakhmanov}. Using the fact that in the TT coordinates $\bar{h}^{TT}_{0\mu}=0$ the linearized Riemann tensor gives \cite{blair}
\begin{equation}
R^{TT}_{i0j0}=-\frac{1}{2c^2}\frac{\partial^2 h^{TT}_{ij}}{\partial \hat{t}^2}+O(h^2)\hskip1mm .\label{riemann}
\end{equation}
This relation is valid for the wave propagation in any direction. collection equations (\ref{diff2})-(\ref{riemann}) together, after neglecting the terms $O(h^2)$, one obtains
\begin{equation}
\frac{d^2\hat{x}^i}{d\hat{t}^2}=\frac{1}{2}\frac{\partial^2 h^{TT}_{ij}}{\partial \hat{t}^2}\hat{x}^j\hskip1mm ,\label{diff3}
\end{equation}
where the second derivative ${\partial^2 h^{TT}_{ij}}/{\partial \hat{t}^2}$ is to be evaluated along the basic geodesic $\hat{x}=\hat{y}=\hat{z}=0$.\\

Imagining that for time $\hat{t}\leqslant 0$ there were no waves $( h^{TT}_{ij}=0)$ in the vicinity of the two observers (at rest with respect to each other), one can write the initial conditions
\begin{equation}
\hat{x}^i (\hat{t})=\hat{x}^i_0=\textsf{constant} \hskip3mm,\hskip3mm \frac{d\hat{x}^i}{d\hat{t}}(\hat{t})=0 \hskip3mm,\hskip3mm \textsf{for}\hskip2mm  \hat{t}\leqslant 0\hskip1mm .\label{initialcond}
\end{equation}
At $\hat{t}=0$ some waves arrives. One expects that $\hat{x}^i(\hat{t})=\hat{x}^i_0+O(h)$ for $\hat{t} > 0$, therefore, because the term $O(h^2)$ is neglected, the equation (\ref{diff3}) can be replaced by
\begin{equation}
\frac{d^2\hat{x}^i}{d\hat{t}^2}=\frac{1}{2}\frac{\partial^2 h^{TT}_{ij}}{\partial \hat{t}^2}\hat{x}^j_0\hskip1mm .\label{diff4}
\end{equation}
Using the initial conditions (\ref{initialcond}), the integration of (\ref{diff4}) gives 
\begin{equation}
\hat{x}^i (\hat{t})=\left( \delta_{ij}+\frac{1}{2} h^{TT}_{ij}(\hat{t}) \right)\hat{x}^i_0\hskip1mm , \hskip3mm \hat{t} > 0 \hskip1mm.\label{almostfinal}
\end{equation}\vskip1mm
\hskip1mm If the spatial axes of the proper reference frame are oriented at such manner that the wave is propagating in the $+\hat{z}$ direction, then 
\begin{equation}
h^{TT}_{ij}=\left[\begin{array}{ccc}
h_{+} & h_{\times} & 0 \\
h_{\times} & -h_{+} & 0 \\
0 & 0 & 0 
\end{array}\right] \cos \left(\omega (\hat{t}-\frac{\hat{z}}{c})\right)\hskip1mm .
\end{equation}
So the equation  (\ref{almostfinal}) can be written in the following explicit form\\
\begin{eqnarray}
\nonumber\hat{x}(\hat{t})&=\hat{x}_0+\frac{1}{2}\left(h_{+}(\hat{t}) \hskip1mm\hat{x}_0+ h_{\times}(\hat{t}) \hskip1mm\hat{y}_0 \right)\hskip1mm,\\[3mm]
\nonumber\hat{y}(\hat{t})&=\hat{y}_0+\frac{1}{2}\left(h_{\times}(\hat{t})\hskip1mm \hat{x}_0+ h_{+}(\hat{t}) \hskip1mm\hat{y}_0 \right)\hskip1mm,\\[3mm]
\hat{z}(\hat{t})&=\hat{z}_0\hskip1mm.
\end{eqnarray}\label{system}
Notice that equations (\ref{system}) indicates that the gravitational wave is transverse. Indeed, it produces relative displacements to the test particles only in the plane perpendicular to the direction of the wave propagation.\\

In this context, the key idea allowing the evaluation of the effect of GW on the laser-interferometer detectors is to consider a perfect ring of point particles initially at rest. Let the radius of the ring be $d_0$ and the center of the ring coincides with the origin of the observer\rq{}s proper reference frame. Then the coordinates of any particle in the ring can be parametrized by using the polar coordinates attached to the center of the ring. Namely :
\begin{equation}
\hat{x}_0=d_0\hskip1mm \cos \phi\hskip1mm,\hskip3mm \hat{y}_0=d_0\hskip1mm \sin \phi\hskip1mm,\hskip3mm \hat{z}_0=0\hskip1mm,\hskip6mm \phi\in[0,2\pi]\hskip1mm.\label{polar}
\end{equation}\vskip1mm
\hskip1mm One can determine the motion of the particles considering the $+$ and $\times$ polarizations separately. If only the $+$ polarization is present, then $h_{\times}=0$ and we have 
\begin{equation}
h^{TT}_{ij}=h_{+}\left[\begin{array}{cc}
1 &  0 \\
0 & -1 
\end{array}\right]\sin\omega \hat{t}\hskip1mm ,
\end{equation}
$\omega$ being the resonant frequency of the gravitational wave\rq{}s source. Then using the equations (\ref{system}) and (\ref{polar}), one gets
\begin{eqnarray}
\nonumber\hat{x}(\hat{t})&=d_0\cos \phi\left(1+\frac{1}{2}h_{+}(\hat{t}) \right)\hskip1mm,\\[3mm]
\hat{y}(\hat{t})&=d_0\sin \phi\left(1-\frac{1}{2}h_{+}(\hat{t}) \right)\hskip1mm.
\end{eqnarray}\label{littlesystem}
It is easy to combine the two equations of (\ref{littlesystem}) to obtain
\begin{equation}
\frac{\hat{x}^2}{\left(a_{+}(\hat{t})\right)^2}+\frac{\hat{y}^2}{\left(b_{+}(\hat{t})\right)^2}=1\hskip1mm,\label{ellipse}
\end{equation}
where
\begin{equation}
a_{+}(\hat{t})=d_0\left(1+\frac{1}{2}h_{+}(\hat{t}) \right)\hskip2mm,\hskip6mm b_{+}(\hat{t})=d_0\left(1-\frac{1}{2}h_{+}(\hat{t}) \right)\hskip1mm.\label{axes}
\end{equation}\vskip1mm\hskip1mm
Equations (\ref{ellipse})-(\ref{axes}) describe an ellipse with the center at the origin of the coordinate system. The ellipse has semi-axes of the lengths $a_{+}(\hat{t})$ and $b_{+}(\hat{t})$, which are respectively parallel to the $\hat{x}$ and $\hat{y}$ axis. $h_{+}(\hat{t})$ being an oscillatory function, the deformation of the initial circle into the ellipse has the following pattern: in the time intervals when $h_{+}(\hat{t}) > 0$, the circle is stretched in the $\hat{x}$ direction and squeezed in the $\hat{y}$ whereas when  $h_{+}(\hat{t}) < 0$, the stretching is along the $\hat{y}$  axis and the squeezing is along the $\hat{x}$ axis.\\

Now let us fix a single particle in the ring. The motion of this particle with respect to the origin of the proper reference frame is given by equations (\ref{littlesystem}), for a fixed value of $\phi$. By eliminating $h_{+}(\hat{t})$, one obtains
\begin{equation}
\frac{\hat{x}^2}{d_0\cos \phi}+\frac{\hat{y}^2}{d_0\sin \phi}-2=0\hskip1mm.
\end{equation}
This means that any single particle in the ring is moving around its initial position along some straight line.\\

If only the cross polarization is present ($h_{+}=0$), on a similar way, one can show 
\begin{eqnarray}
\nonumber\hat{x}(\hat{t})&=d_0\left(\cos \phi+\frac{1}{2}\sin \phi \hskip1mm h_{\times}(\hat{t}) \right)\hskip1mm,\\[3mm]
\hat{y}(\hat{t})&=d_0\left(\sin \phi+\frac{1}{2}\cos \phi \hskip1mmh_{\times}(\hat{t}) \right)\hskip1mm.
\end{eqnarray}\label{crosssystem}

Introducing in the $(\hat{x},\hat{y})$ plane the rotated coordinates $(\hat{x}\rq{},\hat{y}\rq{})$ around the $\hat{z}$ axis by 45 degrees, we have
\begin{equation}
\left[\begin{array}{c} \hat{x}\rq{} \\ \hat{y}\rq{}\end{array}\right]=\left[\begin{array}{cc}
\cos(\pi/4) & \sin(\pi/4) \\
-\sin(\pi/4) & \cos(\pi/4)
\end{array}\right]\left[\begin{array}{c}\hat{x}\\ \hat{y}\end{array}\right]=
\frac{\sqrt{2}}{2}\left[\begin{array}{cc}
1 & 1 \\
-1 & 1
\end{array}\right]\left[\begin{array}{c} \hat{x}\\ \hat{y}\end{array}\right]
\hskip1mm .
\end{equation}
It is easy to rewrite equations (\ref{crosssystem}) in terms of the coordinates $(\hat{x}\rq{},\hat{y}\rq{})$ :
\begin{eqnarray}
\nonumber\hat{x}\rq{}(\hat{t})&=\frac{\sqrt{2}}{2}d_0(\sin \phi+\cos \phi )\left(1+h_{\times}(\hat{t}) \right)\hskip1mm,\\[3mm]
\hat{y}\rq{}(\hat{t})&=\frac{\sqrt{2}}{2}d_0(\sin \phi-\cos \phi )\left(1-h_{\times}(\hat{t}) \right)\hskip1mm.
\end{eqnarray}\label{systemprim}
After eliminating from equations (\ref{systemprim}) the paramiter $\phi$, one gets
\begin{equation}
\frac{\hat{x}\rq{}^2}{\left(a_{\times}(\hat{t})\right)^2}+\frac{\hat{y}\rq{}^2}{\left(b_{\times}(\hat{t})\right)^2}=1\hskip1mm,
\end{equation}
where
\begin{equation}
a_{\times}(\hat{t})=d_0\left(1+\frac{1}{2}h_{\times}(\hat{t}) \right)\hskip2mm,\hskip6mm b_{\times}(\hat{t})=d_0\left(1-\frac{1}{2}h_{\times}(\hat{t}) \right)\hskip1mm.
\end{equation}\vskip1mm\hskip1mm
This means that the initial circle of particles is deformed into an ellipse with its center at the origin of the coordinate system. The ellipse has semi-axes of the lengths $a_{\times}(\hat{t})$ and $b_{\times}(\hat{t})$, which are rotated by an angle of 45 degrees with respect to the $(\hat{x} ,  \hat{y})$-axis.\\

The simplest gravitational wave detector one can imagine, is a body mass $m$ at a distance $L$ from a fiducial laboratory point, connected to it by a spring of resonant frequency $\varpi$ and quality factor $Q$. Einstein equation of geodesic deviation predicts that the infinitesimal displacement $\Delta L$ of the mass along the line of separation from the equilibrium position satisfies in the free falling frame of the observer at the fiducial laboratory frame and for wavelengths $\gg L$ \cite{thorne}
\begin{equation}
\ddot{\Delta L} (t)+2 \frac{\varpi}{Q}\dot{\Delta L}(t)+\varpi^2 \Delta L(t)=\frac{1}{2}\left(F_{+}\ddot{h}_{+}(t)+F_{\times}\ddot{h}_{\times}(t)\right)\hskip1mm,\label{motion}
\end{equation}
where $F_{+}$ and $F_{\times}$ are are coefficients of order unity related to the detector response $h(t)$ to a gravitational wave signal \cite{EE}
\begin{equation}
h(t)=h_{+}(t)\hskip1mm F_{+}+h_{\times}(t)\hskip1mm F_{\times}\hskip1mm,\label{h}
\end{equation}
where
\begin{eqnarray}
F_{+}&=\frac{1}{2}(1+\cos^2\Theta)\cos2\Phi\hskip1mm\cos2\Psi-\cos\Theta\hskip1mm\sin2\Phi\hskip1mm\sin2\Psi\hskip1mm,\\[3mm]
F_{\times}&=\frac{1}{2}(1+\cos^2\Theta)\cos2\Phi\hskip1mm\sin2\Psi+\cos\Theta\hskip1mm\sin2\Phi\hskip1mm\cos2\Psi\hskip1mm,
\end{eqnarray}
$\Theta$, $\Phi$, $\Psi$ being the angles defining the relative orientation of the binary with respect to the detector.\\

Laser-interferometer gravitational wave detectors are composed of two perpendicular km-scale arm cavities with two test-mass mirrors hung by wires at the end of each cavity. The tiny displacement $\Delta L$ of the mirrors induced by a passing gravitational wave are monitored with very high accuracy by measuring the relative optical phase between the light paths in each interferometer arm. The mirrors are pendula with quality factor $Q$ and resonant frequency $\varpi$. Thus, the equation (\ref{motion}) written in Fourier domain, reduces to 
\begin{equation}
\frac{\Delta L}{L}\sim h\hskip1mm.\label{almostfinal}
\end{equation}
\vskip1mm\hskip1mm
In the specific positions we have considered the binary as the source of GW namely, the binary in the ${z}$-direction and the arms of the detector in the ${xy}$-plane, then $\Theta = 0$ \cite{thorne}. Thus $\left|F_{+}\right|$ and $\left|F_{\times}\right|$ which are of order unity, take alternatively the values 0 and 1 on such a way that when one takes the value 0, the other one take the value 1. Now, using (\ref{h}) and taking into account the expressions of $h_{+}$ and $h_{\times}$ (\ref{eq34}), the relation (\ref{almostfinal}) can be written 
\begin{equation}
\left|\frac{\Delta L}{L}\right|\lesssim \left|h_{+}\right|\hskip1mm ,
\end{equation}
which is exactly the expression obtained (\ref{final}) in section 7.
\section{conclusions and discussion}\label{sect:9}\hskip9mm
From the standard GR point of view, the existence of GW is intuitively obvious as soon as the space-time is assumed to act as an elastic medium. The basic properties of GW can then be deduced from the space curvature as a consequence of mass distribution in space and in time.\\
 
Since the pioneering work of Joseph Weber \cite{weber}, the improvement of detectors has been remarkable. Relating them to optical telescopes, the improvement achievement so far is equivalent to the step from a 3 cm diameter optical telescope to a 3 m diameter instrument. In the next decade it is hoped that the improvement will be equivalent to a step up in size from 3 m to 3 Km \cite{ju}. At this sensitivity gravitational wave detection is practically certain, and the field of gravitational astronomy will be able to map and explore the new spectrum, and the objects that it reveals.\\

In this sens, identifying the correct theory of gravity is a crucial issue for modern physics. Also, modifications of gravity are required to deal with the vast phenomenological results. But the ambiguity comes from the fact that very often, these results can be explained successfully both with the standard GR or modified and extended GR. Here we have a blatant example of this situation.\\

In this article, we have adopted an alternative approach to the standard GR by taking into account both curvature and torsion. Including torsion, does not modify the standard approach of GR but bring us out of the Riemann space. In this context, considering a laser interferometer GW detector, we have highlighted the meaning of the breaking of the parallelogram in presence of torsion and, we have shown that the delay time observed in this type of experiments can be interpreted by this approach. Through this delay time, we obtain the usual parameter $\Delta L/L$ which is the relative change of length of the arms of the laser interferometer detector in presence of GW. By considering two specific positions of a typical binary system as the source of GW, namely the NS-NS binary pulsar (1913 +16), we obtain $\log\left|\Delta L/L\right|\lesssim-20.8$ in one position corresponding to a mixture of $+$ and $\times$ polarizations and $\log\left|\Delta L/L\right|\lesssim-21.1$ in the other position corresponding to pure $\times$ polarization.\\
 
It is remarkable that in comparison with the standard GR, not only the numerical results but also the analytical expression, at first order, are exactly the same.\\
  
To go further, it will be  useful to compare these results with those obtained in the context of higher orders gravity. Also, to round off this comparison, it will be necessary to calculate these quantities, adopting the Teleparallel theory of gravity. These are now under study.\\


\section{References}
\begin{thebibliography}{}
\bibitem{ko-no}  S. Kobayashi, K. Nomizu, {\it{Foundations of Differential Geometry.}} Publ. Math. Soc. Japan (1956)
\bibitem{sabata} V. de Sabbata and M. Gasperini, {\it{Introduction to gravitation.}} World Scientific, Singapore (1985)
\bibitem{hehl-maccrea} F. W. Hehl, J. D. Mccrea, E. W. Mielke and  Y. Ne'eman,
{\it{Metric-Affine Gauge Theory of Gravity: Field Equations, Noether Identities, World Spinors, and Breaking of Dilation Invariance}}, Phys.Rept.258, 1 (1995) 
\bibitem{arcos-pereira} H. I. Arcos and J. G. Pereira, {\it{Torsion Gravity: a Reappraisal}}, Int.J.Mod.Phys. D13 (2004)
\bibitem{blanchet} L. Blanchet, Gravitational Radiation from Post-Newtonian Sources and Inspiralling Compact Binaries, Living Rev. Rel. 5, 3 (2002)
\bibitem{hobson} M.~P.~Hobson, G.~P.~Efstathiou and A.~N.~Lasenby, {\it{General Relativity: An introduction for Physicist}}. Cambridge University Press (2006)
\bibitem{Aldrovandi} See f.i. R. Aldrovandi, J. G. Pereira and K. H. Vu, {\it{The Nonlinear Essence of GW}}, Found.Phys.37:1503-1517 (2007)
\bibitem{hehl-vdh} F. W. Hehl, P. von der Heyde, G. D. Kerlick and I. M. Nester, {\it{General Relativity with spin and torsion: Foundations and prospects}}, Rev. Mod. Phys. 48 (1976)
\bibitem{hammond} R. T. Hammond, {\it{Torsion gravity}}, Rep. Prog. Phys. 65. 599 (2002)
 \bibitem{maluf} J. W. Maluf, S. C. Ulhoa and J. F. da Rocha-Neto, {\it{GW and the breaking of parallelograms in space-time}}, General Relativity and Gravitation, 45, 1163 (2013)
 \bibitem{Eisenhart}  L. P. Eisenhart, {\it{An Introduction to Differential Geometry}}. Perinceton University Press (1940)  S. Sternberg, {\it{Lectures on Differential Geometry}}. Perinceton-Hall (1964) M. P. Do Carmo, {\it{Differential Geometry of Curves and Surfaces}}. Perinceton-Hall (1976)
  \bibitem{hough} For more ditails see f. i. 
 M. Pitkin, S. Reid, S. Rowan and J. Hough,  {\it{Gravitational Wave Detection by Interferometry (Ground and Space)}}, Living Rev. Relativity 14, 5 (2011)
 \bibitem{ju} L. Ju, D.G. Blair and C. Zhao, {\it{Detection of GW}},  Rep. Prog. Phys. 63 (2000) 1317-1427
 \bibitem{verbunt} F. Verbunt, {\it{Waiting for LISA: binaries with orbital periods less than
$\sim10^4 s$}}, Class. Quantum Grav. 14 (1997) 1417–1423
\bibitem{weber} J. Weber,{\it{Detection and generetaion of gravitational waves}}, Phys. Rev. 117 (1960)
\bibitem{misner} C. Misner, K.S. Thorne, and J.A. Wheeler, {\it{Gravitation}} W.H. Freeman and Company, NewYork (1973)
\bibitem{rakhmanov} M. Rakhamanov, {\it{Response of test masses to GW in the local Lorentz gauge}} Phys. Rev. D 71, 084003 (2005)
\bibitem{blair} D.G. Blair {\it{The Detection of GW}}, Cambridge University Press (1991)
\bibitem{thorne} K.S. Thorne, {\it{Gravitational radiation, in 300 Years of Gravitation}}, Cambridge University Press (1987)
\bibitem{EE} E.E. Flanagan and S.A. Hughes, {\it{ Measuring GW from binary black hole coalescences. I. Signal to noise for inspiral, merger, and ringdown}} Phys. Rev. D 57, 4535 (1998)
\bibitem{ligo} A. Abramovici et al., {\it{LIGO - The Laser Interferometer Gravitational-Wave Observatory}}, Science 256, 325 (1992); http://www.ligo.org
\bibitem{virgo} B. Caron and al., {\it{ The Virgo Interferometer}} Class. Quant. Grav. 14, 1461 (1997); http://www.virgo.infn.it
\bibitem{geo}H. Lück and the GEO600 team, {\it{  The geo-600 project}} Class. Quant. Grav. 14, 1471 (1997); http://www.geo600.uni-hannover.de
\bibitem{tama}M. Ando and al., {\it{ the TAMA collaboration}} Phys. Rev. Lett. 86, 3950 (2001); http://tamago.mtk.nao.ac.jp
\end{thebibliography}
\end{document}